\documentclass[aps,prl,twocolumn]{revtex4}
\usepackage{times}
\usepackage{graphicx}

\begin{document}
\title{Are structural biases at protein termini a signature of
vectorial folding?} \author{Alessandro Laio$^1$ and Cristian
Micheletti$^2$\\ $\null$\\ \small$^1$ Computational Science, Dept. of
Chemistry and Applied Biosciences, ETH Zurich. \\ \small c/o USI
Campus, Via Buffi 13, CH-6900 Lugano, Switzerland\\ \small$^2$
International School for Advanced Studies (SISSA) and INFM, Via Beirut
2-4, 34014 Trieste, Italy}

\begin{abstract}
Experimental investigations of the biosynthesis of a number of
proteins have pointed out that part of the native structure can be
acquired already during translation. We carried out a comprehensive
statistical analysis of some average structural properties of proteins
that have been put forward as possible signatures of this progressive
buildup process. Contrary to a widespread belief, it is found that
there is no major propensity of the amino acids to form contacts with
residues that are closer to the N terminus. Moreover, it is found that
the C terminus is significantly more compact and locally-organized
than the N one. Also this bias, though, is unlikely to be related to
vectorial effects, since it correlates with subtle differences in the
primary sequence. These findings indicate that even if proteins aquire
their structure vectorially no signature of this seems to be
detectable in their average structural properties.
\end{abstract}

\maketitle

\section{Introduction}

\vskip 1.0cm Anfinsen's principle states that the folded state of an
isolated protein corresponds to the \emph{global} minimum of the
system free energy at physiological temperature
\cite{anfinsenscheraga}. The Anfinsen's view is supported by the
experimentally observed reversibility of the folding process for a
large class of proteins. It appears so well-established that it
provides the conceptual starting point of most theoretical studies and
{\em ab initio} computational simulations of protein folding
\cite{Funnel1,Funnel3,Funnel5,Pande98,Duan98,topomer99,topomerPlaxco,rosetta}.
Indeed, computational approaches based on molecular dynamics or
stochastic sampling rely on the notion that, independently of the
starting unfolded conformation, the interplay of amino acid
interactions is sufficient to drive a protein to the global free
energy minimum.

The validity of the Anfinsen's principle appears surprising
considering that the native structure of a protein is the result of a
complicated mechanism which starts with the ribosomal translation, may
involve the action of molecular chaperons and may end in
post-translational modifications. The experimental investigation of
the biosynthesis of specific proteins has led to the formulation of
the cotranslational
hypothesis\cite{fedorov95,komar97,fedorov97,kolb01}. According to it,
proteins which fold {\em in vivo} acquire their spatial structure in
the course of translation through specific kinetic routes in which the
already grown peptide influences the folding of the rest of the chain.
It should be remarked that cotranslational folding is not necessarily
in contradiction with Anfinsen's principle. Indeed, virtually all the
experimental and theoretical investigation of cotranslational folding
state explicitly that the same final (native) conformations are
achieved as a result of the biosynthesis or of the refolding from the
denatures state. Despite this observation, several putative native
structural signatures of the progressive build-up of nascent proteins
have been put forward over the years, ranging from the absence of
knots in folded proteins, to the atypical proximity of the two
termini. Some of these signatures have later been shown to be void of
statistical significance \cite{baldwin96}. To the present day, a
feature that is still invoked in favour of the progressive quenching
of nascent proteins into their native structure is the different
structural organisation of the two termini. Since the N-terminal
region is the first to exit from the ribosomal tunnel, it is expected
to be more locally organized and compact than the C-terminal region
which should grow over the pre-formed protein scaffold
\cite{schulz88,alexandrov93,chakra02,fedorov97}. This stimulating
suggestion followed the observation that the conformation of
N-terminal regions appeared to be easier to predict than the C-
counterparts \cite{schulz88}. More recently, Alexandrov analyzed a
collection of protein conformations with the aim of detecting
signatures of vectorial growth \cite{alexandrov93}. The study
concluded that in about two thirds of the analyzed proteins the
majority of residues formed more contacts with amino acids that
preceded rather than followed them in the primary sequence. This was
interpreted as a clear signature of the progressive structural
build-up propagating from the N terminus. According to
ref.~\cite{alexandrov93} this asymmetry would imply that, typically,
the N-terminal part of the protein is more compact than the C-terminal
one, since ``previous'' contacts in the N-terminal region are, by
necessity, local. The latter suggestion was, however, not supported by
the comparison of the termini in terms of common and intuitive
measures of compactness.

In this study we re-examine these issues and several others related to
the structural inequality of the N and C termini. We find that,
according to several definitions of compactness, it is the C terminus
that is more compact than the N one, in contradiction with the result
of ref.~\cite{alexandrov93}. The different bias in compactness is
shown to originate from a larger propensity of the C terminus to
attain helical conformations. To clarify whether the observed
inequality is compatible with the thermodynamic hypothesis we
elucidate its relationship with the difference in amino acidic
composition and sequence of the two termini. The observed structural
bias appears to be encoded in the primary sequences, in agreement with
Anfinsen's principle.

\section{Methods}

The structures on which the analysis is performed were obtained
starting from the PDBselect list of about 2000 non-homologous proteins
in the protein data bank \cite{pdbselect2,pdb}. As customary
\cite{chakra02,GORIV,thornton82}, of these structures we retained only
those comprising at least 80 amino acids, without incomplete or
ambiguous structural information, and not containing signal peptides.
The identification of putative signal peptides (for either prokaryotes
or eukaryotes) was carried out using the approach of
ref.\cite{signalpeptides} based on two different types of neural
networks. Proteins that had a probability greater than 50\% to contain
signal peptides according to both methods were removed from the
set. The selection procedure resulted in the 373 monomeric proteins
and 85 multi domain proteins listed in Table \ref{table:proteins}.

The results discussed in the following sections are obtained through a
statistical analysis performed on the monomeric proteins only, but are
practically unchanged if the multi domain proteins are included in the
data set.

Each protein was analyzed to detect structural biases at the two
termini and to trace their possible origin back to the primary
sequence. Several structural measures were used to characterise the
average properties of terminal segments of increasing lengths at the N
and C ends. Part of the analysis is carried out in terms of the
contact matrix, $\Delta$. For each protein, the contact matrix
element, $\Delta_{l,m}$ reflects the spatial proximity of the $l$th and
$m$th residues in the protein. Denoting with $d_{lm}$ the distance of
the corresponding C$_{\alpha}$ atoms (taken as interaction centroids
for the whole residues) in the native structures, the strength of the
contact interaction is calculated from the sigmoidal weight:
$\Delta_{l,m}= [1-\tanh(d_{lm}-R_c)]/2$. As customary, the cutoff
interaction, $R_c$, was set to 7.5 \AA.

The contact matrix is used to detect the possible preferential
directions along the primary sequence of the contacts between amino
acids. In the same spirit of Alexandrov we computed the average
fraction of previous contacts for each residues, $r_p$. In terms of
the contact matrix, $r_p$ is defined as $r_p(i)=\sum_{j <i}
\Delta_{i,j} / \sum_{j \not= i} \Delta_{i,j}$ (note that nearest
neighbors are included in the sum). The previous/forward character of
each residues is then assigned according to whether $r_p$ is greater
or smaller than 0.5. It is important to notice that the asymmetry of
the previous/forward character both at the level of site and of the
whole protein is not in contradiction with the symmetry of the contact
matrix, $\Delta$.

We also considered the average number of contacts, $n_c(i)$, that
amino acids at a given sequence distance, $i$, from the nearest
terminus make with residues that are closer, along the primary
sequence, to the same terminus. The definition of $n_c(i)$ for the
N-terminus is $n_c(i-1) = \langle \sum_{j < i} \Delta_{i,j} \rangle$
while for the C-terminus case is $n_c(i) = \langle \sum_{j > (L-i)}
\Delta_{L-i,j} \rangle$. In these formulae, $L$ is the length of the
protein under consideration and the brackets denote the average over
the proteins in the data set; also consecutive residues are excluded
from the summation. Since the data set is built from a non-redundant
set of $N_p$ proteins ($N_p$ = 373 and 85 respectively for the
monomeric and multimeric ones) the statistical uncertainty on $n_c(i)$
is calculated as $\sigma_i/\sqrt{N_p}$ where $\sigma^2_i$ is the
second moment of the number of contacts at distance $i$ from the N (or
C) terminus. The statistical significance of the difference in the
value of $n_c(i)$ observed at the N and C termini is finally
calculated using the Students t-test.

As further measures of compactness of the termini we also considered
the radius of gyration, $R_G(i)$ of the segments stretching up to the
$i$th residue from the N or C termini as well as the fraction of local
contacts. To correlate the observed structural inequality at the two
termini with biases in the primary sequence we also considered a
number of sequence-based observables as a function of the distance $i$
from the nearest terminus. In particular we considered
\begin{itemize}
\item[(a)] the average hydrophobicity according to the Kyte-Doolittle
scale :Ala=1.8; Cys=2.5 ; Leu=3.8; Met=1.9; Glu=-3.5; Gln=-3.5;
His=-3.2; Lys=-3.9; Val=4.2; Ile=4.5; Phe=2.8; Tyr=-1.3; Trp=-0.9;
Thr=-0.7; Gly=-0.4; Ser=-0.8; Asp=-3.5; Asn=-3.5; Pro=-1.6;
Arg=-4.5.
\item[(b)] the average steric hindrance defined as the total number
of heavy atoms (not hydrogens) in the side chain of an amino acid.
\item[(c)] the average helical content assigned according to the DSSP
algorithm. The helical character of a residue is set equal to 1 if it is
classified as H (alpha helix) G (3/10 helix) or I (pi helix), and 0
otherwise.
\end{itemize}

Besides these observables we have also considered the helical
propensities predicted by the GOR-IV algorithm described in
ref. \cite{GORIV}. In this method the probability of an amino acid to
belong to an alpha helix is estimated from its primary-sequence
neighborhood, through a set of coefficients expressing the conditional
probability that a given pair of amino acids at a fixed sequence
separation belongs to a secondary structure motif. These coefficients
are learned on the set of 373 selected proteins from which we removed
all the residues at a sequence separation smaller than 30 from each of
the two termini. By doing so we ensure the statistical reliability of
the GOR-IV results for the proteins' termini, since none of the
structural motifs to be predicted is included in the training set.
The original source code of the GOR IV program was compiled setting
the Nterm and Cterm parameters equal to zero, so to allow secondary
structure predictions also for residues very close to the terminus
(otherwise set to "coil" by default). Suitable normalisation factors
of the knowledge-based weights were also introduced to account for the
fact that the averaging window can span less than the default number
of 17 residues if the site is at a sequence distance smaller than 9
from either termini.

\section{Results and discussion}

In Ref. \cite{alexandrov93} it is suggested that amino acids have a
higher propensity to form contacts with residues that are closer to
the N terminus. This was interpreted as a signature of the progressive
structural buildup propagating from the N terminus.

\begin{figure}[h!]
\includegraphics[width=3.2in]{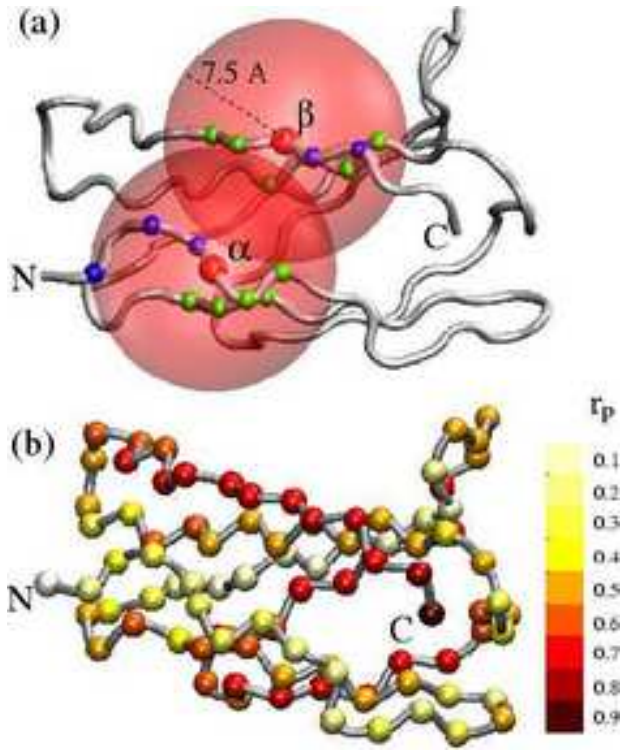}
\caption{Backbone of protein 1ttg. (a): Previous and forward contacts
for the two residues, $\alpha$ and $\beta$, at sequence separation 6
from the N and C-termini (highlighted in red). The spheres denote the
region of space within an interaction cutoff distance of 7.5 \AA\ from
the two reference residues. Amino acids within this cutoff distance
and with a smaller [longer] sequence separation than the reference
residue from the nearest terminus are highlighted in blue
[green]. (b): Color-coded profile for the fraction of ``previous''
contacts, $r_p$, for each site of 1ttg. Accordingly, for the reference
sites of the top panel, we have $r_p(\alpha) \approx 3/8 $
and $r_p(\beta) \approx 6/8$.}
\label{fig:prevforw}
\end{figure}

 The previous/forward bias originally observed by Alexandrov
\cite{alexandrov93} is confirmed by the analysis of our data set,
though with notable changes in perspective and conclusions. In order
to quantitatively characterise the bias, we here compute the average
fraction of previous contacts for each residue, $r_p$, see
Fig. \ref{fig:prevforw}b. It is found that $r_{p}$ is rather
independent on the length of the proteins in the data set and is
practically unaffected by the omission of residues at the protein's
termini. In terms of the sequence separation, the autocorrelation
length in the values of $r_p$ is about 4. The average value of $r_{p}$
in our set is $0.504 \pm 0.002$ where the statistical error on the
mean was calculated from the dispersion of the sample and accounting
for the sequence-separation correlation. If one assigns the previous
[forward] identity to individual amino acids based on the fact that
$r_p$ is greater [smaller] than 0.5, one finds that, of the nearly
40,000 residues, 50.6 \% of them are of type previous. The asymmetry
in the directional preference of contact formation therefore appears
to be minimal. This tiny asymmetry is amplified by the procedure of
ref.~\cite{alexandrov93} where a hierarchy of majority rules was used
to assign a previous/forward character first to residues and then to
proteins. In fact, the site-wise assignment of the previous/forward
character can be used to define the character of ``blocks'' of
consecutive residues according to the majority rule. One therefore
finds that, for (non-overlapping) blocks of size 5, 11 and 17
residues, the fraction of previous-type blocks is 51.6, 52.3 and
54.1\%, respectively. It is therefore clear that the majority rule
amplifies the slight site-wise asymmetry in a manner that is dependent
on the block-size. Consequently, the heterogeneity of proteins lengths
in a data set make problematic the proper notion of the average
previous/forward character of proteins. Even in this case, however,
the procedure ref.~\cite{alexandrov93} applied to our data set yields
a fraction of proteins of type ``previous'' equal to 59\%, a number
substantially lower than the 75\% observed in ref
~\cite{alexandrov93}. In summary, the previous/forward asymmetry,
though statistically well-founded, appears to be much smaller than
originally stated. This bias, previously regarded as a signal of the
N-terminal initiation and propagation of the folding process, may
possibly reflects the genuine chemical inequality of peptide chains
under inversion of the primary sequence \cite{inverse_similarity03}.

\section{Structural differences between the termini}

Another possible signature of the progressive build-up of the proteins
is that, since the N-terminal region is the first to exit from the
ribosome, it is expected to be organized differently than the
C-terminal region which should grow over the pre-formed protein
scaffold \cite{schulz88,alexandrov93,chakra02,fedorov97}. To elucidate
this, we carried out a detailed analysis of structural differences
between proteins' termini.

\begin{figure}[h!]
\includegraphics[width=3.2in]{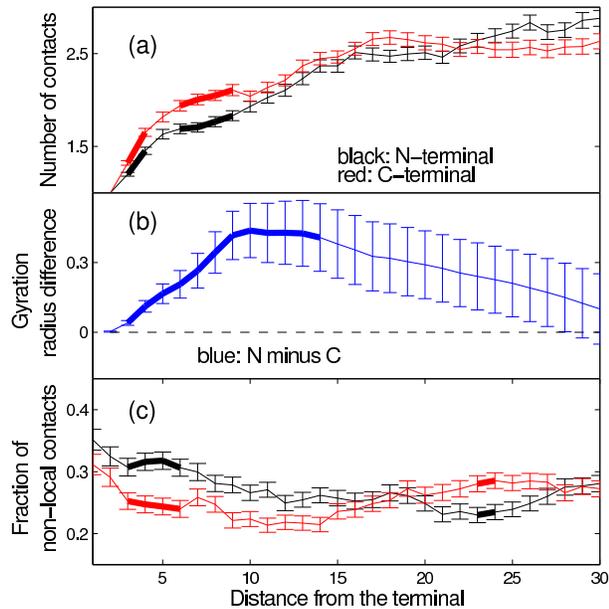}
\caption{Average values of structural observables as a function of the
distance $i$ of the residue from the nearest terminus. Averages are
taken over the 373 structures listed in Table I. The error bars are
the standard deviation on the average. The thick lines correspond to
regions where the statistical significance (Students t-test) of the
N-C pointwise difference is larger than 99 \%. (a): Average number of
contacts that amino acids at a given sequence distance from the
nearest terminal make with residues that are closer to the same
terminal. (b): Average difference (in \AA) between the gyration
radius of segments of increasing length at the N and C termini. (c):
Fraction of non-local contacts, i.e. interactions with residues at a
sequence separation larger than 6.}
\label{fig:fig2}
\end{figure}

We first consider the average number of contacts, $n_c(i)$, that amino
acids at a given sequence distance, $i$, from the nearest terminus
make with residues that are closer, along the primary sequence, to the
same terminus. The widespread notion that the N-terminus is more
compact than the C one\cite{schulz88,alexandrov93,chakra02,fedorov97},
and the tiny previous/forward bias we observed, would imply that
$n_c(i)$ should be higher for the N region. As visible in
Fig.~\ref{fig:fig2}a, however, the observed bias contradicts this
expectation. In a region that extends up to 10 residues away from the
termini it is the C region that appears to be richer in internal
contacts by an amount that has a high statistical significance. The
difference is still larger than the error bar at a distance of 20 from
the termini. The conclusions are robust against changes of the
interaction cutoff, $R_c$, in the viable range of 6--8 \AA\, and upon
the use of a step function instead of a sigmoidal one for weighting
the interactions. It is important to remark that $n_c(i)$ reflects a
propensity to form contacts {\em within} the terminal regions,
i.e. disregarding interactions with residues with sequence distance
greater than $i$ from the reference terminus. In fact, if one
considers the contacts made with any residue irrespective of the
sequence separation from the terminus, then no statistical difference
between the two termini emerge.

To clarify the structural basis for the bias shown in
Fig.~\ref{fig:fig2}a we monitored the average gyration radius of the
terminal regions. This quantity is a direct measure of the difference
in compactness. The results, shown in Fig.~\ref{fig:fig2}b,
demonstrate that the C-terminal region has a smaller average
radius. For instance, the average gyration radius of the first and the
last 15 amino acids of a protein are 9.1 and 8.7 \AA\
respectively. The difference has a statistical relevance higher than
99\% up to $i=16$. Finally, we analyzed the average propensity of the
amino acids to form non-local contacts, i.e. contacts with residues at
a sequence separation larger than 6 \cite{topomerPlaxco}. Also in
this case there is a statistically-significant difference revealing a
greater propensity of the C-terminal region to form local contacts
than the N-terminal counterpart, as visible in Fig.~\ref{fig:fig2}c.
These results unambiguously show that, on average, the C-terminus is
more compact than the N-one.

\begin{figure}[h!]
\includegraphics[width=3.2in]{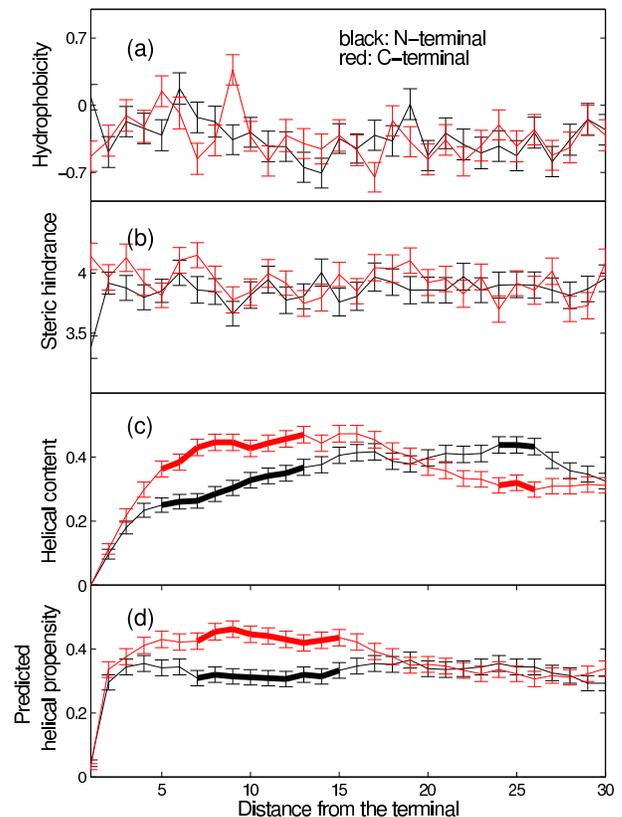}
\caption{ Average values of sequence-based observables as a function
of the distance, $i$, of the residue from the nearest terminus. Line
colors, thickness and error bars follow the same convention of Figure
\ref{fig:fig2}. (a): Average hydrophobicity according to the
Kyte-Doolittle scale (b): Average steric hindrance, (c): Average
structural helical content identified with the DSSP algorithm. (d):
Average helical propensity as predicted by the GOR-IV algorithm.}
\label{fig:fig3}
\end{figure}

Since a low contact order in proteins is an indicator of high helical
content\cite{Plaxco98} we have analyzed the secondary content in the
proteins' terminal regions by means of the DSSP
algorithm\cite{DSSP}. The results, shown in Fig.~\ref{fig:fig3}c,
highlight the higher probability of the C-terminal region to attain
helical conformations, up to at least $i=15$, consistently with other
structural studies \cite{thornton82,secondary_biased}. As apparent in
Fig.~\ref{fig:fig3}c, the helical content has a maximum at separation
$i \approx 15$ for the C terminus and at separation $i \approx 25$ for
the N one. As a result, for $ 20 < i < 30$, it is instead more common
to observe helices in the N-terminal regions. This
secondary-structure bias is responsible not only for the observed
difference of contact order but also for the higher number of contacts
formed within the C-terminal region rather than the N one: since the
average overall number of total contacts is the same in the two
regions, a higher helical content implies a higher number of local
contacts.

The observed differences between the two termini are highly
statistically significant, and their relationship to vectorial growth
must be addressed, since the presence of structural biases could be an
evidence in favor of conformational trapping resulting from an
out-of-equilibrium build up \cite{chakra02,alexandrov93}. However, an
average difference in the structure of the termini is not necessarily
in contradiction with Anfinsen's hypothesis. In fact, the structural
bias may reflect a systematic difference in the primary sequence at
the termini. Such differences in sequence composition have already
been reported in the literature \cite{berezovsky99} but have been
found to be statistically significant only for residues very close to
the termini, and it would be surprising if these small differences
would result in the bias we observed.

In order to quantitatively address this issue, we considered several
sequence-related properties, looking for an average difference between
the N and the C regions. Remarkably, the two termini could not be
distinguished by any \emph{point-wise} (i.e. single-amino acid)
property we considered. An example is provided in
Fig.~\ref{fig:fig3}a, where we plot the average hydrophobicity as a
function of the distance from the termini: no statistically
significant difference is observed between the two regions. We also
considered the average helical propensity, estimated through the
knowledge-based Chou-Fasman\cite{chou} parameters, and the average
steric hindrance, defined as the number of side chain heavy atoms
(Fig.~\ref{fig:fig3}b). Also for these indicators we did not observe
any statistically significant difference of the two termini. We have
also tested if the two termini are distinguished by the effective
pairwise interactions \cite{KGS,Miyazawa96} among amino
acids. Considering residues up to a given separation from the each
terminus we have calculated the energy resulting from the interaction
of all pairs of residues in the segment. By doing so we ascertain if,
within the limitations of the energy scoring function, the two termini
have different average propensities for self-interaction, and hence
compactness or locality. Also in this case no
statistically-significant difference was found.

These results do not necessarily imply that the termini structural
inequality is not encoded by the primary sequence, since it may
only reflect the limitations of point-wise and mean-field indicators.
To improve the analysis we resorted to the powerful GOR-IV scheme
\cite{GORIV} for predicting proteins' secondary structure from the
mere knowledge of their sequence. The information theory approach of
ref.~\cite{GORIV} was chosen because it has a good prediction
performance and yet does not rely on structural alignment, which could
bias the prediction at the termini.

The results are summarized in Fig.~\ref{fig:fig3}d and reveal that the
GOR-IV approach is able to predict the correct bias on the termini
helical propensity, at least in the $i<15$ region. The same
conclusions hold using a jackknife scheme were the prediction on each
protein is done with the parameters learnt on all other proteins in
the set.  The N- and C-terminal difference in the average helical
content predicted by the GOR IV algorithm is of the same order of the
structurally-observed one. While the average behaviour is thus
captured, on individual proteins the typical fraction of residues in
helical conformations that are correctly-predicted is about $p_1=$
0.65, while the fraction of non-helical residues that are mistakenly
predicted as helical is $p_2=$0.17. These number may be taken as
indicators of the average reliability of the predictions. Over our
finite sample we saw that $p_1$ and $p_2$ are equal to 0.62 and 0.16
respectively over residues at distance 5--17 from the N-terminus and
0.66 and 0.22 for the C-terminal ones. Though these fluctuations may
simply reflect the finiteness of the sample, it is instructive to
consider them as genuine differences in the performance of the GOR IV
scheme at the termini. Even in this case the putative difference in
performance would be responsible for less than half the difference in
predicted helical propensity. According to the Students t-test, the
remaining pointwise difference would have a probability of less than
7\% to be generated by chance. We observe that the region of high
statistical significance in Fig. ~\ref{fig:fig3}d, spans many more
residues that those over which GOR predictions appear to be correlated
(4.5 residues).

This shows that the difference in helical content and,
therefore, the difference in compactness between the termini, is
indeed encoded in the primary sequence, though it cannot be picked up
by intuitive point-wise indicators \cite{berezovsky99}.

\section{Conclusions}

In order to elucidate the possible role of out-of-equilibrium effects
in determining the native structures of proteins, we analyzed the
structural differences between the two termini. We have found that the
C terminus has a higher helical content (Fig.~\ref{fig:fig3}a), a
smaller gyration radius (Fig.~\ref{fig:fig2}b), and contains a larger
number of local contacts (Fig.~\ref{fig:fig2}a and
Fig.~\ref{fig:fig2}c) than the N terminus. These results contradict
previous observations that, based on the intuitive image of a
progressive protein build up, argued for the higher compactness of the
N terminus. The use of a sequence-based secondary structure prediction
method revealed that the observed structural asymmetry of the termini
is encoded in subtle difference of the primary sequence at the protein
ends. This is consistent with the Anfinsen's hypothesis while it rules
out the necessity to invoke out-of-equilibrium effects to account for
the terminal structural inequality. Of course, the possibility
that naturally-selected proteins have evolved so to exploit kinetic
biases to reach the global free energy minimum cannot be ruled out, as
already envisaged by Levinthal
\cite{Levinthal,Goconsistent,minimalfrustr,Funnel3}. The results
presented here pose the question of the biophysical rationale behind
the sequence and structural inequality of the two termini. Though
this issue is beyond the scope of the present analysis, it is tempting
to speculate that the presence of this average terminal difference
across a large set of unrelated proteins may be the result of
evolutionary pressure, e.g. for folding cooperativity \cite{chan}.

{\bf Acknowledgments}
We are indebted to S. Alberti for stimulating discussions on nascent
proteins and to J.-F. Gibrat for providing us the GOR-IV program. We
are also thankful to R. Bulo, H-S. Chan, G. Colombo, F. Gervasio,
E. Lattmann M. Parrinello and S. Raugei, for several useful comments
and discussions.

\begin{table}[h!]
\begin{center}
\begin{tabular}{| l l l l l l l l l l |}
\hline
1jaj & 1ib8 & 1l6u & 1n9j & 1l7y & 1f53 & 1n9d & 1wtu & 1wkt & 1lab \\
1n91 & 1f40 & 1gnc & 1f16 & 1j5i & 1j57 & 1clh & 1ckv & 1lmz & 1j3g \\
1ttg & 1j2o & 1n3k & 1n3j & 1cj5 & 1tnn & 1tiu & 1bsh & 1qkl & 1buy \\
1sxl & 1bvh & 1bw3 & 1m4o & 1m4p & 1eo1 & 1emw & 1iqo & 1h6q & 1ej5 \\
1eio & 1irl & 1i4v & 1cdb & 1rip & 1mwb & 1mm4 & 1qr5 & 1ddb & 1df3 \\
1jr5 & 1jrm & 1g7o & 1jt8 & 1jw3 & 1g4g & 1jyt & 1nzt & 1k0h & 1nwb \\
1jjj & 1hqi & 1orm & 1fzt & 1ab3 & 1gh8 & 1nr3 & 1adn & 1k8h & 1ghh \\
1ag4 & 1jfw & 1d1r & 1pba & 1ap7 & 1aps & 1jdq & 1ily & 1fo5 & 1fjc \\
1kot & 1cur & 1b6f & 1hce & 1fbr & 1b9r & 1yub & 1o3s & 1fez & 1f5t \\
1exj & 1qle & 1fq1 & 1b26 & 1maw & 1rpt & 1qpv & 1gtp & 2avi & 1jum \\
1im0 & 2ldx & 1wbc & 1l9a & 1oft & 2nmt & 168l & 1qso & 1b9l & 7mht \\
1j5s & 1jfm & 1odg & 1nlx & 1cid & 1mok & 3pva & 1jik & 4ald & 1ysc \\
1ki9 & 1j0c & 1ulb & 1by3 & 1i9b & 1fvp & 1ci0 & 1ib1 & 1exc & 1ef9 \\
1jr4 & 1dk4 & 1iw7 & 1d9u & 1lwh & 1g0t & 1hcn & 1efp & 1i4w & 1ltb \\
1dir & 1kdq & 1g5z & 1cc5 & 1qhh & 1ixy & 1oo5 & 1h21 & 1fb3 & 2thi \\
1hup & 1hm8 & 1i4n & 1ith & 1g1a & 1lpb & 1ib5 & 1kfq & 1mow & 1b35 \\
1pdy & 1h3q & 1eoi & 1kho & 1cvm & 2sas & 1gsq & 2pfk & 1k0k & 1ggl \\
1jgs & 1bl0 & 1k3r & 1ej3 & 1g71 & 1n5d & 1ee6 & 1itq & 8prn & 1kgt \\
1m1b & 1gan & 1a65 & 1l5p & 1ufh & 1eje & 1i8n & 1fvz & 2mjp & 1jlx \\
1iof & 1kte & 1jyb & 1jzk & 1ji3 & 1e5f & 1ko9 & 1k3b & 1kvs & 1ash \\
1id2 & 1i4z & 1g64 & 1otg & 1eom & 1i0i & 1hzi & 1eum & 1dm9 & 1oa9 \\
1el6 & 1bys & 1hbk & 1coz & 1ipb & 1it6 & 1tl2 & 2ubp & 1cpm & 1k04 \\
1qhd & 1ig3 & 1jhs & 1mr8 & 1c2a & 1qmy & 1b93 & 1b8a & 1ogh & 1gak \\
1jmv & 1jku & 1ew2 & 1g8e & 1fpo & 1dqe & 2spc & 1hdk & 1mug & 1mml \\
1cmb & 1jq3 & 1jyh & 1rpj & 1jh6 & 1ld8 & 1e1a & 1hxn & 1c44 & 1pgs \\
1iab & 1mqv & 1dk0 & 1cv8 & 1kpt & 1gwy & 1gxy & 1qst & 1mk4 & 1dqi \\
2bop & 1kzq & 1htw & 1nep & 1ecs & 1thx & 1mol & 1ako & 1l2q & 1d3v \\
1cqm & 1cxy & 1iby & 1cnv & 1dj7 & 1hqk & 1it2 & 1gvp & 1jhj & 1fi2 \\
1b5e & 1fs7 & 1uaq & 3pvi & 1lyc & 1na3 & 1iv3 & 1cip & 1i0r & 1p1m \\
1g2q & 1lmi & 1bx4 & 1whi & 1f7l & 1l7a & 1n7o & 1opd & 1ezm & 1dqz \\
1lo7 & 1dj0 & 1nf9 & 1brt & 1bqc & 2sns & 8abp & 1dzk & 1uca & 1lc5 \\
1jig & 1qre & 1idp & 1is3 & 1aba & 1e6u & 1fp2 & 1es5 & 3vub & 1ew4 \\
1eca & 4eug & 1jl7 & 1iw0 & 1oaf & 1ezg & 1llf & 1h2w & 1n8v & 2lis \\
1jl1 & 1es9 & 1o8x & 1dbf & 1flm & 1lq9 & 1ka1 & 1qdd & 1qks & 1qau \\
1j96 & 1ird & 1e29 & 1obo & 1jg1 & 1o08 & 1m15 & 1gu2 & 1h97 & 1m1n \\
1i8o & 1kng & 1n8k & 1jf8 & 1k7c & 1kt6 & 1oh0 & 1qlw & 1f86 & 1ql0 \\
1qj4 & 1c5e & 1n62 & 1jcl & 1m2d & 1psr & 1kqp & 1mfm & 1c7k & 1ga6 \\
1ug6 & 1k4i & 1iqz &   &   &   &    &      &      &      \\
\hline\hline
1tic & 1qrj & 1ffk & 1ffk & 1ffk & 1ffk & 1ffk & 1ffk & 1ffk & 1m57 \\
2atc & 1fl7 & 1qle & 1iis & 1f51 & 1n32 & 1n32 & 1n32 & 1n32 & 1n32 \\
1qb3 & 1gph & 1hm7 & 1nbq & 1prt & 1prt & 1fvv & 1nsk & 1i50 & 1k83 \\
1i50 & 1is7 & 1gvm & 1qax & 1khr & 1mae & 1kf6 & 1kf6 & 1kf6 & 1i78 \\
1iw7 & 1lhr & 1bvp & 1h31 & 1f5q & 1pdn & 1bmq & 1qhh & 1qhh & 1nbw \\
1jrk & 1jj2 & 1jtd & 1mbx & 1jc5 & 1epb & 1cz3 & 1h4m & 1k8k & 1cew \\
1kx5 & 1gyh & 1hke & 1n71 & 1lmb & 1jiw & 1lj9 & 1o26 & 1gy7 & 1gpq \\
4ubp & 1lk9 & 1gd0 & 1o9r & 1fm0 & 1o7n & 1gk8 & 1jo0 & 1i0d & 1hyo \\
1mqk & 1mqk & 1qft & 2tps & 1m1n &      &      &      &      &      \\
\hline
\end{tabular} 
\end{center}
\caption{PDB codes of the 373 monomeric proteins (top) and the 85
multimeric proteins (bottom) used for the sequence/structure
analysis.}
\label{table:proteins}
\end{table}


\begin{thebibliography}{10}

\bibitem{anfinsenscheraga}
Anfinsen, C. and Scheraga, H.~A.
\newblock Experimental and theoretical aspects of protein folding.
\newblock Adv. Protein Chem. 29:205--299, 1975.

\bibitem{Funnel1}
Bryngelson, J.~D., Onuchic, J.~N., Socci, N.~D., and Wolynes, P.~G.
\newblock Funnels and pathways and the energy landscape of protein folding: {A}
  synthesis.
\newblock Proteins 21:167--195, 1995.

\bibitem{Funnel3}
Dill, K.~A. and Chan, H.~S.
\newblock From {Levinthal} to pathways to funnels.
\newblock Nature Struct. Biol. 4:10--19, 1997.

\bibitem{Funnel5}
Sali, A., Shakhnovich, E., and Karplus, M.
\newblock How does a protein fold.
\newblock Nature 369:248--251, 1994.

\bibitem{Pande98}
Pande, V.~S., Grosberg, A.~Y., Tanaka, T., and Rokshar, D.~S.
\newblock Pathways for protein folding: is a new view needed.
\newblock Curr. Opin. in Struct. Biol. 8:68--79, 1998.

\bibitem{Duan98}
Duan, Y. and {Kollman}, P.~A.
\newblock Pathways to a protein folding intermediate observed in a
  1-microsecond simulation in aqueous solution.
\newblock Science 282:740--749, 1998.

\bibitem{topomer99}
Debe, D.~A., Carlson, M.~J., and Goddard, W.~A.
\newblock The topomer-sampling model of protein folding.
\newblock Proc. Natl. Acad. Sci. USA 96:2596--2601, 1999.

\bibitem{topomerPlaxco}
Makarov, D.~E. and Plaxco, K.~W.
\newblock The topomer search model: a simple, quantitative theory of two-state
  protein folding kinetics.
\newblock Prot. Sci. 12:17--26, 2003.

\bibitem{rosetta}
Rohl, C.~A., Strauss, C.~E., Misura, K.~M., and Baker, D.
\newblock Protein structure prediction using {Rosetta}.
\newblock Method. Enzymol. 383:66--93, 2004.

\bibitem{fedorov95}
Fedorov, A.~N. and Baldwin, T.~O.
\newblock Contribution of cotranslational folding to the rate of formation of
  native protein structure.
\newblock Proc. Natl. Acad. Sci. USA 92:1227--1231, 1995.

\bibitem{komar97}
Komar, A.~K., Kommer, A., Krasheninnikov, I.~A., and Spirin, A.~S.
\newblock Cotranslational folding of globine.
\newblock J. Biol. Chem. 272:10646--10651, 1997.

\bibitem{fedorov97}
Fedorov, A.~N. and Baldwin, T.~O.
\newblock Cotranslational protein folding.
\newblock J. Biol. Chem. 272:32715--32718, 1997.

\bibitem{kolb01}
Kolb, V.~A.
\newblock Cotranslational protein folding.
\newblock Mol. Biol. 35:584--590, 2001.

\bibitem{baldwin96}
Christopher, J.~A. and Baldwin, T.~O.
\newblock Implications of the {N} and {C}-terminal proximity for protein
  folding.
\newblock J. Mol. Biol. 257:175--187, 1996.

\bibitem{schulz88}
Schulz, G.~E.
\newblock A critical evaluation of methods for prediction of protein secondary
  structures.
\newblock Ann. Rev. Biophys. Biophys. Chem. 17:1--21, 1988.

\bibitem{alexandrov93}
Alexandrov, N.
\newblock Structural argument for {N}-terminal initiation of protein folding.
\newblock Protein Science 2:1989--1991, 1993.

\bibitem{chakra02}
Bhattacharyya, R., Pal, D., and Chakrabarti, P.
\newblock Secondary structures at polypeptide-chain termini and their features.
\newblock Acta Cryst. D 58:1793--1802, 2002.

\bibitem{pdbselect2}
Hobohm, U. and Sander, C.
\newblock Enlarged representative set of protein structures.
\newblock Prot. Sci. 2:522, 1992.

\bibitem{pdb}
Bernstein, F.~C., Koetzle, T.~F., Williams, G.~J., Meyer, E.~E., Brice, M.~D.,
  Rodgers, J.~R., Kennard, O., Shimanouchi, T., and Tasumi, M.
\newblock The {Protein} {Data} {Bank}: a computer-based archival file for
  macromolecular structures.
\newblock J. Mol. Biol. 112:535--542, 1977.

\bibitem{GORIV}
Garnier, J., Gibrat, J.~F., and Robson, B.
\newblock {GOR} method for predicting the secondary structure from amino acid
  sequence.
\newblock Method. Enzymol. 266:540--553, 1996.

\bibitem{thornton82}
Thornton, J.~M. and Chakauya, B.~L.
\newblock Conformation of terminal regions in proteins.
\newblock Nature 298:294--297, 1982.

\bibitem{signalpeptides}
Bendtsen, J.~D., Nielsen, H., {von Heijne}, G., and Brunak, S.
\newblock Improved prediction of signal peptides: Signalp 3. 0.
\newblock J. Mol. Biol. 340:783--795, 2004.

\bibitem{inverse_similarity03}
Lorenzen, S., Gille, C., Preissner, R., and Fr\"ommel, C.
\newblock Inverse sequence similarity of proteins does not imply structural
  similarity.
\newblock FEBS Lett. 545:105--109, 2001.

\bibitem{Plaxco98}
Plaxco, K.~W., Simons, K.~T., and Baker, D.
\newblock Contact order, transition state placement and the refolding rates of
  single domain proteins.
\newblock J. Mol. Biol. 277:985--994, 1998.

\bibitem{DSSP}
Kabsch, W. and Sander, C.
\newblock Dictionary of protein secondary structure: pattern recognition of
  hydrogen-bonded and geometrical features.
\newblock Biopolymers 22:2577--2637, 1983.

\bibitem{secondary_biased}
Bhattacharyya, R., Pal, D., and Chakrabarti, P.
\newblock Secondary structures at polypeptide-chain termini and their features.
\newblock Acta Cryst. 58:1793--1802, 2002.

\bibitem{berezovsky99}
Berezovsky, I.~N., Kilosanidze, G.~T., Tumanyan, V.~G., and Kisselev, L.~L.
\newblock Amino acid composition of protein termini are biased in different
  manners.
\newblock Prot. Engin. 12:23--30, 1999.

\bibitem{chou}
Chou, P.~Y. and Fasman, G.~D.
\newblock Prediction of secondary structure.
\newblock Adv. Enzymol. 47:45--148, 1978.

\bibitem{KGS}
Kolinsky, A., Godzik, A., and Skolnick, J.
\newblock A general method for the prediction of the three dimensional
  structure and folding pathway of globular proteins: Application to designed
  helical proteins.
\newblock J. Chem. Phys. 98:7420--7433, 1993.

\bibitem{Miyazawa96}
Miyazawa, S. and Jernigan, R.~L.
\newblock Residue-residue potentials with a favorable contact pair term an
  unfavorable high packing density term, for simulation and threading.
\newblock J. Mol. Biol. 256:623--644, 1999.

\bibitem{Levinthal}
Levinthal, C.
\newblock Mossbauer Spectroscopy in Biological Systems.
\newblock Univ. Illinois Press, , 1969.

\bibitem{Goconsistent}
Go, N.
\newblock Theoretical studies of protein folding.
\newblock Ann. Rev. Biophys. Bioeng. 12:183--210, 1983.

\bibitem{minimalfrustr}
Wolynes, P.~G.
\newblock Symmetry and the energy landscapes of biomolecules.
\newblock Proc. Natl. Acad. Sci. USA 93:14249--14255, 1996.

\bibitem{chan}
Chan, H.~S., Shimizu, S., and Kaya, H.
\newblock Cooperativity principles in protein folding.
\newblock Method. Enzymol. 380:350--379, 2004.

\end{thebibliography}
\end{document}